\documentclass[showpacs,preprintnumbers,floatfix,twocolumn]{revtex4}

\usepackage{graphicx}\begin{document}
\title{Resonant and rolling droplet}

\author{S. Dorbolo, D. Terwagne, N. Vandewalle, and T. Gilet}
\affiliation{GRASP, Physics Department B5, University of Li\`{e}ge, B-4000 Li\`{e}ge, Belgium }

\begin{abstract}When an oil droplet is placed on a quiescent oil bath, it eventually collapses into the bath due to gravity. The resulting
coalescence may be eliminated when the bath is vertically vibrated. The droplet bounces periodically on the bath, and the air layer between the
droplet and the bath is replenished at each bounce. This sustained bouncing motion is achieved when the forcing acceleration is higher than a
threshold value. When the droplet has a sufficiently low viscosity, it significantly deforms : spherical
harmonic \boldmath{$Y_{\ell}^m$} modes are excited, resulting in resonant effects on the threshold acceleration curve. Indeed, a lower acceleration is
needed when $\ell$ modes with $m=0$ are excited. Modes $m \ne 0$ are found to decrease the bouncing ability of the droplet.
When the mode $\ell=2$ and $m=1$ is excited, the droplet rolls on the vibrated surface without touching it, leading to a new self-propulsion
mode.\end{abstract}

\maketitle

Bouncing droplets have been first studied by Couder {\it et al} \cite{Couder:2005}. With a radius $R < 0.5$~mm, they may generate local waves on
a 50~cSt bath and use them to move horizontally and interact with their surroundings~: they detect submarine obstacles, they orbit together or
form lattices, they are diffracted when they pass through a slit \cite{Couder:2005b,Protiere:2006,Couder:2006,Vandewalle:2006,Lieber:2007}.
Bouncing droplets exhibit a wide variety of motions and interactions that present some strong analogies with quantum physics, astronomy and
statistical physics.  

\begin{figure} [t]
\includegraphics[width=6cm]{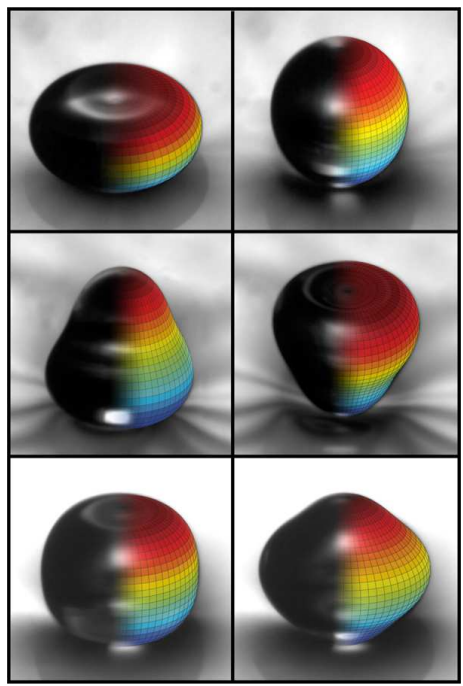}
\caption{\label{fig:Modesm0} {\bf Various deformation modes of a bouncing droplet (\boldmath{$\nu=1.5$}~cSt, $R=0.765$~mm) observed with a high-speed
camera for various forcing frequencies}. The first line (resp. 2nd, 3rd) displays a mode $Y_2^0$ (resp. $Y_3^0$, $Y_4^0$) with $m=0$
(axisymmetry). The forcing frequency is 50~Hz (resp. 160~Hz, 275~Hz) and the reduced acceleration $\Gamma = 0.3$ (resp. 2, 6). First and second
columns represent snapshots at two different times of the oscillation. The spherical harmonic solution (on the right of each picture) is
superposed to the experimental pictures (on the left of each picture).}
\end{figure}

In contrast with these previous studied, we choose a bath viscosity of 1000~cSt, a droplet viscosity between 1.5 and
100~cSt and a droplet radius $R$ between 0.76 and 0.93~mm. With a bath viscosity at least 10 times larger as the droplet viscosity, the bath
deformations are inhibited (capillary waves are fully damped) while the harmonic deformations of the droplet are enhanced. As shown recently in
\cite{Gilet:2008}, the droplet deformation ensures its bouncing ability. Various modes of deformation may be excited as depicted in
Fig.\ref{fig:Modesm0}.  Each picture has been constructed from an experimental snapshot of the droplet (the left side of each picture) and the
calculated 3D spherical harmonic (the right side).  Those modes are analogous to the natural modes of deformation introduced by Rayleigh
\cite{Rayleigh:1879} and may be expressed in terms of spherical harmonics $Y_{\ell}^m$. The simplest mode that may be used for bouncing is the
mode $Y_2^0$ \cite{Gilet:2008}.  We will show that droplets can use non-axisymetric mode $Y_{2}^1$ to move horizontally, or more precisely to
roll over the bath.  This new mode of self-propulsion drastically contrasts with the bouncing walker mode described in \cite{Protiere:2006}
since the bath deformations are not necessary for the propulsion.

On a bath vertically vibrated according to a sinusoidal motion $A \sin(2\pi f t)$, the periodic bouncing only occurs when the reduced maximal
vertical acceleration of the bath $\Gamma = 4 \pi^2 A f^2 / g$ is higher than a threshold value $\Gamma_{th}$, where $g$ is the acceleration of
gravity. Formally, $\Gamma_{th}$ depends on the forcing frequency $f$ \cite{Gilet:2008}, the droplet radius $R$ \cite{GiletT:2007}, and the
physical parameters of the liquids (density $\rho$, viscosity $\nu$, surface tension $\sigma$). We measure the threshold $\Gamma_{th}$ as a
function of the forcing frequency for various droplet sizes and viscosities. For each frequency, we place a droplet on the bath in a bouncing
configuration, i.e. $\Gamma > \Gamma_{th}$. The forcing acceleration is then decreased progressively. When the threshold is reached, the droplet
cannot sustain anymore the periodic bouncing and quickly coalesces with the bath. Since the droplet and the bath are made from different
liquids, a coalescence event locally contaminates the bath. Therefore, droplets always need to be placed at different locations on the bath.

\begin{figure} [htbp]
\includegraphics[width=\columnwidth]{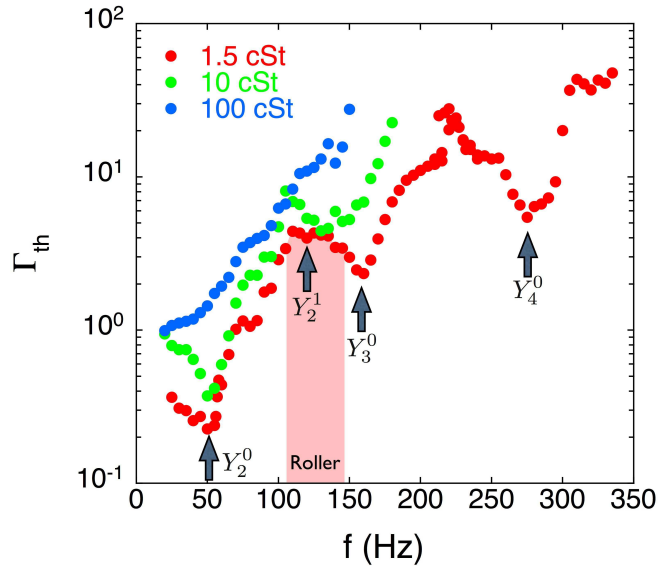}
\caption{\label{fig:Gammaf} {\bf Evolution of the bouncing threshold acceleration \boldmath{$\Gamma_{th}$} with respect to the forcing frequency.} The red,
green and blue bullets correspond to droplets with viscosity $\nu=$1.5, 10 and 100~cSt respectively, and with a constant radius $R=0.765$~mm.
Depending on the forcing frequency, various modes are observed, that may be related to spherical harmonics $Y_{\ell}^m$. }
\end{figure}

In Fig.\ref{fig:Gammaf}, the threshold acceleration $\Gamma_{th}$ is represented as a function of $f$, for droplets with viscosities $\nu=$1.5,
10 and 100~cSt. The threshold acceleration for the 100 cSt droplet increases with the frequency in a monotone way.  By opposition, the 1.5~cSt curve is characterized by regularly spaced local minima that, in some way, correspond to a resonance of the system :
a minimal energy supply is required to sustain the periodic bouncing motion. The first minimum, at $f=50$~Hz, corresponds to $\Gamma_{th}=$0.25,
a value significantly less than the 1~g minimal threshold required by inelastic bouncing objects on a vibrated plate. Modes $m=0$ and $\ell =$2,
3 and 4 are observed for forcing frequencies corresponding to minima in threshold acceleration curve. The higher the frequency, the higher-order
the excited mode, and the higher the corresponding threshold acceleration. We note that the asymmetric mode $Y_2^1$ occurs at a local maximum of
the threshold curve.

The bouncing droplet may be considered as an oscillating system analogous to the damped driven harmonic oscillator : surface tension is the
restoring force and viscosity the damping process. The dimensionless ratio between both is the Ohnesorge number $Oh = \nu \sqrt{\rho} /
\sqrt{\sigma R}$, which is equal to 0.012, 0.078 and 0.775 for a droplet viscosity of 1.5, 10 and 100~cSt respectively. When $Oh \ll 1$, the
viscous damping is negligible and resonance is important. Damping increases as $Oh$ gets closer to 1. As seen in Fig.\ref{fig:Gammaf}, the
bouncing droplet reacts to an increased damping in the same way as the harmonic oscillator : (\textit{i}) the resonance frequency slightly
decreases, shifting the whole threshold curve to the left, (\textit{ii}) the required input $\Gamma_{th}$ increases, and (\textit{iii}) extrema
tend to disappear. At 100~cSt ($Oh=0.775$), the damping is fully active and the threshold curve increases monotonically with the frequency : no
more resonance is observed. This high-viscosity behaviour has already been observed by Couder \textit{et al} \cite{Couder:2005} for 500~cSt
droplets bouncing on a 500~cSt bath. Those authors proposed to model the threshold curve by $\Gamma_{th}=1+\alpha f^2$ where $\alpha$ depends on
the droplet size among others. This equation is obtained by only considering the motion of the mass center of the droplet and the squeezing of
the air film between the droplet and the bath, without considering the droplet deformation. This model only fits high viscous regimes in
contrast with the model developed in \cite{Gilet:2008}.

\begin{figure} [htbp]
\includegraphics[width=\columnwidth]{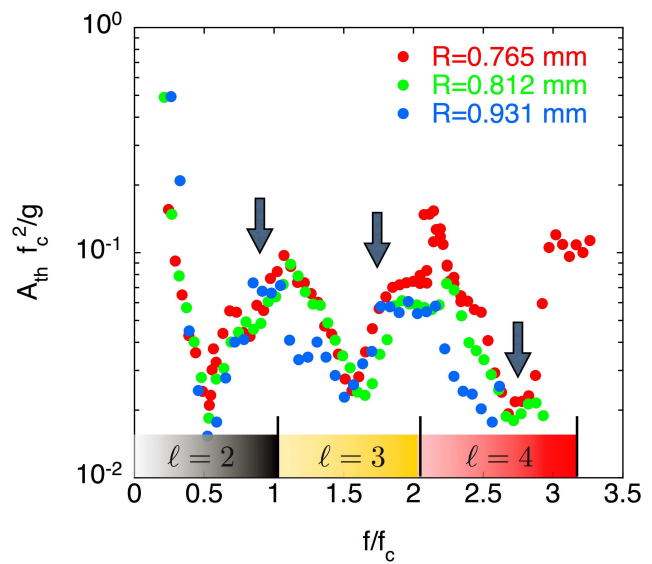}
\caption{\label{fig:Af} {\bf Dimensionless amplitude \boldmath{$A_{th} f_c^2/g$} as a function of the reduced forcing frequency $f/f_c$}, where $A_{th}$ is the
amplitude of the bath vertical motion corresponding to the reduced acceleration $\Gamma_{th}$ and $f_c=\sqrt{M/\sigma} f$ the capillary
frequency of a droplet of mass $M=4\pi/3 \rho R^3$. Blue, green and red bullets correspond to droplet radius $R=0.765$~mm, 0.812~mm and 0.931~mm
respectively. The droplet viscosity is $\nu=1.5$~cSt. Modes $\ell=2$, $\ell=3$ and $\ell=4$ are observed in the black, yellow and red ranges of
frequency. Boundaries between those zones correspond to maxima in the threshold curve. Moreover, those particular frequencies may be obtained by
multiplying the Rayleigh natural frequencies (Eq.\ref{eq:Rayleigh}) by a factor 1.15. Arrows indicate the Rayleigh frequencies, which cannot be
directly related to inflections in the threshold curve.}
\end{figure}

In 1879, Lord Rayleigh \cite{Rayleigh:1879} described natural oscillations of an inviscid droplet. Since those oscillations are due to surface
tension, natural frequencies scale as the capillary frequency $f_c = \sqrt{\sigma/M}$, where $M=4\pi/3 \rho R^3$ is the droplet mass. More
exactly, the dispersion relation prescribes the natural "Rayleigh" frequency $f_R$ related to a $\ell$-mode:
\begin{equation} \label{eq:Rayleigh}
\biggl(\frac{f_R(\ell)}{f_c}\biggr)^2=F(\ell) = \frac{1}{3\pi} \ell (\ell-1) (\ell+2)
\end{equation}
The function $F(\ell)$ may vary (frequencies are shifted by a multiplicative factor), depending on the way the droplet is excited \cite{Courty:2006,Noblin}. For free oscillations,
Eq.(\ref{eq:Rayleigh}) is degenerated according to the $m$ parameter. In Fig.~\ref{fig:Af}, threshold data obtained for various droplet sizes
collapse on a single curve by using the Rayleigh scaling. Moreover, at the bouncing threshold, the vertical force resulting from the droplet
deformation exactly balances the gravity. One may define a characteristic length $L=g/f_c^2$ corresponding to the free fall distance during the
capillary time $1/f_c$. As shown in Fig.~\ref{fig:Af}, the threshold amplitude $A_{th} = \Gamma_{th} g /(4\pi^2 f^2)$ scales as the length $L$,
whatever the droplet size. The minimum value of $A_{th} f_c^2/g$ does not vary significantly with the mode index $\ell$. In Fig.~\ref{fig:Af},
the natural frequencies defined in Eq.(\ref{eq:Rayleigh}) and represented by arrows do correspond neither to the minima in threshold, nor to the
maxima. However, these frequencies multiplied by 1.15 give the maxima positions, at $f/f_c = 1.05, 2.05 \mbox{ and } 3.15$ for $\ell=$2, 3 and 4
respectively. This numerical factor depends on the geometry of the excitation mode (bouncing in this case) : e.g. another (smaller) factor is
obtained when the droplet is stuck on a vibrated solid surface \cite{Courty:2006,Noblin}.

As shown in \cite{Gilet:2008}, the bouncing ability of droplets is due to the cooperation of (i) the droplet deformation, that stores potential
energy, and (ii) the vertical force resulting from the squeezing of the intervening air layer between the droplet and the bath. The forced
motion of the bath provides some energy to the droplet, a part of which helps the droplet to bounce (translational energy) while the other part
increases internal motions inside the droplet, which are eventually dissipated by viscosity. The proportion of energy supplied to the
translational / internal motion varies with the forcing frequency (i.e. when the energy is provided in the oscillation cycle). Minima (resp.
maxima) in the threshold curve correspond to a maximum (resp. minimum) of the translational to internal energy ratio. Maxima may be related to
the cut-off frequency recently observed and theoretically explained by Gilet \textit{et al.} \cite{Gilet:2008}. This fact is confirmed
experimentally, since the maxima in threshold correspond to the boundaries between modes. At those points, corresponding to a forcing frequency
of $1.15\; f_R$, the most energy is spent in internal motions : the droplet resonates and absorbs energy.

\begin{figure} [htbp]
\includegraphics[width=6cm]{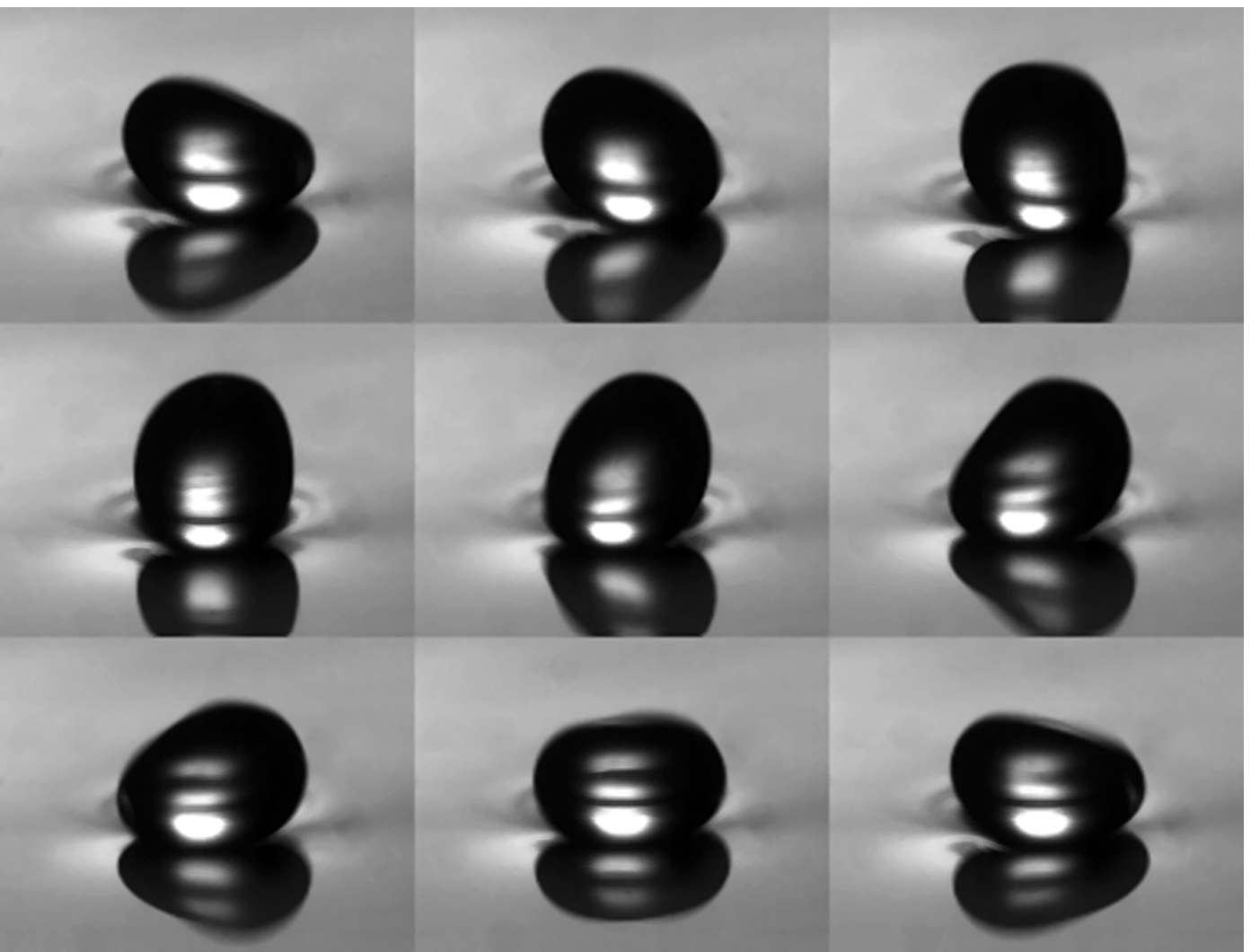}
\includegraphics[width=6cm]{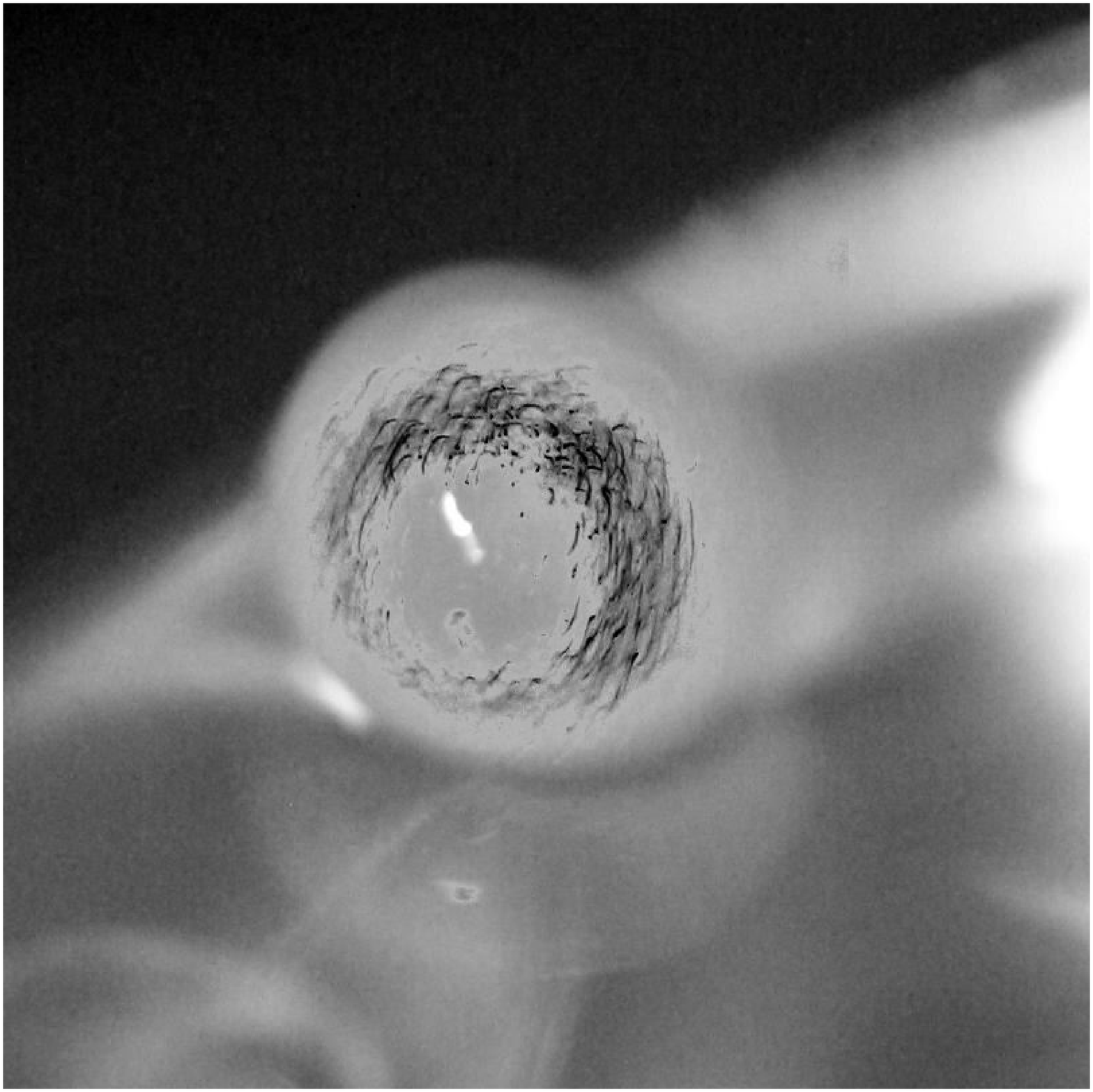}
\caption{{\bf Rolling motion of the droplet} (top) Mode $Y_2^1$ of a bouncing droplet ($\nu=1.5$~cSt, $R=0.765$~mm) observed at $f=115$
Hz and $\Gamma= 4.5 \geq \Gamma_{th}$. Frames are separated by 1~ms. The droplet is rolling towards the left. (bottom) The rolling motion is
revealed by small bright particles spread inside the droplet. }
\end{figure}

The droplet behaviour has been observed in the vicinity of the first maximum in the threshold curve. At 115 Hz, the droplet moves along a linear
trajectory analogously as the walkers observed in \cite{Couder:2005b}. In this latter case, 50 cSt droplets were produced on a 50 cSt silicon oil
bath. The mechanism for the walking motion is the interaction between the droplet and the bath surface wave generated by the bouncing. Such a
mechanism does not hold in the case of a 1000 cSt silicon oil bath. Indeed, the generated waves are rapidly damped and cannot be responsible for
the motion of the droplet. A movie of the moving droplet has been recorded using a high speed camera (Fig. 5). The images are separated by 1 ms.
The deformation is not axi-symmetrical as in the resonant minima of the threshold acceleration curve.  On the other hand, the mode is related to
$\ell=2$ and $m=1$. This mode is characterized by two lines of nodes that are orthogonal.  It results the existence of two fixed points located
on the equator of the droplet. As the the line of nodes does not follow the axi-symmetrical geometry, those fixed points move. More precisely,
they turn giving the droplet a straight direction motion. As far as Rayleigh frequencies are concerned, we observe a break of degeneracy for the
$m$ parameter!

\begin{figure} 
\includegraphics[width=\columnwidth]{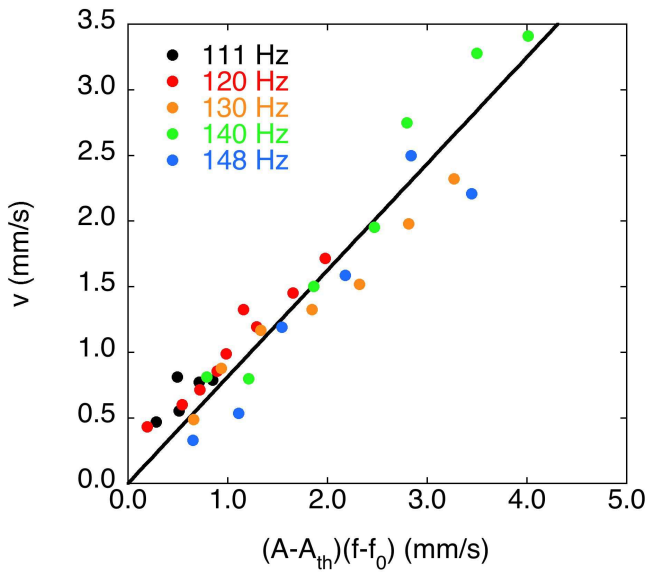}
\caption{\label{fig:speed} {\bf Initial speed of the roller for various frequencies (indicated in the legend) as a function of the characteristic speed \boldmath{$v^*$}}.}
\end{figure}

Some tracers have been placed in the droplet. The reflection of the light on the particles allows to follow the inner fluid motion, which
clearly reveals that the droplet rolls over the bath surface.  The initial speed $v$ of the droplet has been measured in respect to the forcing
parameters (amplitude and frequency).  A scaling is found when considering that the phenomenon can occur only above a cut-off frequency $f_0
\approx 103$ Hz and the amplitude threshold $A_{th} (f)=\Gamma_{th}(f) g/ (4 \pi^2 f^2)$ as
\begin{equation}
v=2 \pi \alpha (A-A_{th}(f))(f-f_0)
\end{equation}
where $\alpha$ is a constant. The initial speed is represented versus the characteristic speed $v^*=(A-A_{th}(f))(f-f_0)$ in
Fig.\ref{fig:speed}. The proportionality between the initial speed of the roller and $v^*$ is remarkable.
Three characteristic frequencies are involved in the rolling process : (i) the excitation time 100 Hz, (ii) the rotation of the liquid inside
the droplet about 10 Hz and (iii) the translational motion about 0.2~Hz (the droplet travels a distance corresponding to its circumference once
per 5 seconds). The three processes are interrelated, i.e. the deformation induces the internal motion which induces the rolling and the
translational motion.

The deformation of low viscosity bouncing droplets is emphasized on a high viscosity bath : the droplet oscillations are much less damped than
the bath oscillations. Depending on the forcing frequency, droplets need a different amount of supplied energy to achieve a sustained periodic
bouncing. When the forcing frequency corresponds to a multiple of the eigenfrequency of the bouncing droplet ($f \sim (\ell-1) f_c$), the
supplied energy is mainly lost into internal motions, so the threshold acceleration $\Gamma_{th}$ is maximum. On the other hand, $\Gamma_{th}$
is minimum when $f/f_c=\ell-3/2$.

A new self-propelled mode has been discovered for forcing frequencies between the $\ell=2$ and $\ell=3$ modes (between roughly 100 and 150 Hz)
for a sufficiently low droplet viscosity. This excited mode is a non-axi-symmetrical mode  $Y_2^1$ characterized by an internal rotation of the
fluid inside the droplet.  The droplet rolls over the vibrated bath ! This droplet displacement technique is more adapted to large low-viscosity
droplets; it is therefore of considerable interest for microfluidic applications: manipulating aqueous mixtures without touching them.

SD/TG would like to thank FNRS/FRIA for financial support.   The Authors want to thank also COST P21 Action 'Physics of droplets' (ESF) for
financial help.  J.P. Lecomte  (Dow Corning, Seneffe, Belgium) is thank for silicon oils.  A. Kudrolli (Clark University, MA, USA), J.Y. Raty
(University of Li\`ege), P. Brunet (University of Lille, France) and R. D'Hulst are acknowledged for fruitful discussions.


\begin{thebibliography}{99}
\bibitem{Couder:2005} Y. Couder, E. Fort, C. H. Gautier and A. Boudaoud, From bouncing to floating: noncoalescence of drops on a fluid bath, Phys. Rev. Lett. {\bf 94}, 177801 (2005)

\bibitem{Couder:2005b} Y. Couder, S. Proti\`{e}re, E. Fort and A. Boudaoud, Walking and orbiting bouncing droplets, Nature {\bf 437}, 208 (2005)

\bibitem{Protiere:2006} S. Proti\`{e}re, A. Boudaoud and Y. Couder, Particle-wave association on a fluid interface, J. Fluid Mech. {\bf 554}, 85 (2006)

\bibitem{Couder:2006} Y. Couder and E. Fort, Single-particle diffraction and interference at a macroscopic scale, Phys. Rev. Lett. {\bf 97}, 154101 (2006)

\bibitem{Lieber:2007} S. Lieber, M. Hendershott, A. Pattanaporkratana and J. Maclenna, Phys. Rev. E {\bf 75}, 056308 (2007)

\bibitem{Vandewalle:2006} N. Vandewalle, D. Terwagne, K. Mulleners, T. Gilet and S. Dorbolo, Dancing droplets onto liquid surfaces, Phys. Fluids {\bf 18}, 091106 (2006)

\bibitem{GiletT:2007} T. Gilet,  N. Vandewalle and S. Dorbolo, Controlling the partial coalescence of a droplet on a vertically vibrated bath, Phys. Rev. E {\bf 76}, 035302 (2007)

\bibitem{Gilet:2008} T. Gilet, D. Terwagne, N. Vandewalle and S. Dorbolo, Dynamics of a bouncing droplet onto a vertically vibrated interface, Phys. Rev. Lett. {\bf 100}, 167802 (2008)

\bibitem{Landau:1959} L. Landau and E. Lifchitz, Fluid Mechanics (vol. 6) (Addison Wesley, 1959)

\bibitem{Rayleigh:1879} Lord Rayleigh, Proc. R. Soc. London {\bf 29}, 71 (1879)

\bibitem{Courty:2006} S. Courty, G. Lagubeau and T. Tixier, Phys. Rev. E {\bf 73}, 045301(R) (2006)

\bibitem{Noblin} X. Noblin, 1. Buguin, and F. Brochard-Wyart, Vibrated sessile drops: Transition between pinned and mobile contact line oscillations, Europhys. J. E {\bf 14}, 395-404 (2004)

\end{thebibliography}
\end{document}